\documentclass[%
notitlepage,
reprint,
superscriptaddress,
 amsmath,amssymb,
 aps,
prb,
floatfix,
]{revtex4-2}
\usepackage{graphicx}
\usepackage{dcolumn}
\usepackage{bm}
\usepackage{hyperref}
\hypersetup{
  colorlinks=true,
  linkcolor=blue,
  filecolor=magenta,
  urlcolor=cyan,
  citecolor=green,
  pdfpagemode=FullScreeen
}
\usepackage{booktabs}
\usepackage[table,xcdraw]{xcolor}
\usepackage{mathtools}

\newcommand{\beginsupplement}{%
        \setcounter{table}{0}
        \renewcommand{\thetable}{S\arabic{table}}%
        \setcounter{figure}{0}
        \renewcommand{\thefigure}{S\arabic{figure}}%
     }

\begin{document}


\title{Evidence of extreme domain wall speeds under ultrafast optical excitation}

\author{Rahul Jangid}
\affiliation{Department of Materials Science and Engineering, University of California Davis, Davis, CA, USA}%

\author{Nanna Zhou Hagstr{\"o}m}%
\affiliation{Department of Physics, Stockholm University, 106 91 Stockholm, Sweden}%
\affiliation{Department of Materials Science and Engineering, University of California Davis, Davis, CA, USA}%

\author{Meera Madhavi}
\affiliation{Department of Materials Science and Engineering, University of California Davis, Davis, CA, USA}%

\author{Kyle Rockwell}
\affiliation{Center for Magnetism and Magnetic Nanostructures, University of Colorado Colorado Springs, Colorado Springs, CO, USA}%

\author{Justin M. Shaw}
\affiliation{Quantum Electromagnetics Division, National Institute of Standards and Technology, Boulder, CO, USA}%

\author{Jeffrey A. Brock}
\affiliation{Center for Memory and Recording Research, University of California San Diego, La Jolla, CA, USA}%

\author{Matteo Pancaldi}
\affiliation{Elettra Sincrotrone Trieste S.C.p.A., Area Science Park, S.S. 14 km 163.5, 34149 Trieste, Italy}%

\author{Dario De Angelis}
\affiliation{Elettra Sincrotrone Trieste S.C.p.A., Area Science Park, S.S. 14 km 163.5, 34149 Trieste, Italy}%

\author{Flavio Capotondi}
\affiliation{Elettra Sincrotrone Trieste S.C.p.A., Area Science Park, S.S. 14 km 163.5, 34149 Trieste, Italy}%

\author{Emanuele Pedersoli}
\affiliation{Elettra Sincrotrone Trieste S.C.p.A., Area Science Park, S.S. 14 km 163.5, 34149 Trieste, Italy}%

\author{Hans T. Nembach}
\affiliation{Department of Physics, University of Colorado, Boulder, Colorado 80309, USA}
\affiliation{Associate, Physical Measurement Laboratory, National Institute of Standards and Technology, Boulder, Colorado 80305, USA}%

\author{Mark W. Keller}
\affiliation{Quantum Electromagnetics Division, National Institute of Standards and Technology, Boulder, CO, USA}%

\author{Stefano Bonetti}
\affiliation{Department of Physics, Stockholm University, 106 91 Stockholm, Sweden}%
\affiliation{Department of Molecular Sciences and Nanosystems, Ca’ Foscari University of Venice, 30172 Venezia, Italy}%

\author{Eric E. Fullerton}
\affiliation{Center for Memory and Recording Research, University of California San Diego, La Jolla, CA, USA}%

\author{Ezio Iacocca}
\affiliation{Center for Magnetism and Magnetic Nanostructures, University of Colorado Colorado Springs, Colorado Springs, CO, USA}.

\author{Roopali Kukreja}
\affiliation{Department of Materials Science and Engineering, University of California Davis, Davis, CA, USA}%

\author{Thomas J. Silva}
\affiliation{Quantum Electromagnetics Division, National Institute of Standards and Technology, Boulder, CO, USA}%

\date{April 27, 2023}

\begin{abstract}
Time-resolved ultrafast EUV magnetic scattering was used to test a recent prediction of $>$10 km/s domain wall speeds by optically exciting a magnetic sample with a nanoscale labyrinthine domain pattern. Ultrafast distortion of the diffraction pattern was observed at markedly different timescales compared to the magnetization quenching. The diffraction pattern distortion shows a threshold-dependence with laser fluence, not seen for magnetization quenching, consistent with a picture of domain wall motion with pinning sites. Supported by simulations, we show that a speed of $\approx$ 66 km/s for highly curved domain walls can explain the experimental data. While our data agree with the prediction of extreme, non-equilibrium wall speeds locally, it differs from the details of the theory, suggesting that additional mechanisms are required to fully understand these effects.

\end{abstract}
\maketitle
The ability to manipulate mesoscopic-scale magnetization \cite{heydermanMesoscopicMagneticSystems2021} has potential applications in ultra-low power magnetic memory and logic \cite{parkinMagneticDomainWallRacetrack2008, carettaFastCurrentdrivenDomain2018, manchonCurrentinducedSpinorbitTorques2019}. For example, current-driven domain wall speeds greater than 5 km/s have been demonstrated with bilayers composed of a compensated ferrimagnet and Pt \cite{carettaFastCurrentdrivenDomain2018}. Exceeding these current-driven domain wall speeds is dependent either on future material breakthroughs or developing novel routes for controlling magnetic behavior. Far-from-equilibrium physics \cite{prigogineTimeStructureFluctuations1993, hemmingerDirectingMatterEnergy2007} in ultrafast conditions \cite{tvetenElectronmagnonScatteringMagnetic2015, hellmanInterfaceinducedPhenomenaMagnetism2017, durrXRayViewUltrafast2016} offer a unique possibility due to the introduction of novel dissipative pathways that are not accessible under equilibrium. In fact, a recent theoretical study by \citet{balazDomainWallDynamics2020} predicts that extremely fast domain wall speeds of $\approx$14 km/s in ferromagnets can be achieved via optical pumping due to superdiffusive spin currents \cite{battiatoSuperdiffusiveSpinTransport2010}. This is a remarkable prediction as it exceeds the generally accepted maximum speed for ferromagnets of $\approx$100 m/s for domain walls. Domain walls, which can be considered as bound magnetic solitons (localized nonlinear excitations with finite energy) \cite{Kosevich1990}, undergo Walker-breakdown above these speeds and the soliton-like structure of a domain wall becomes unstable \cite{schryerMotion180Domain1974, ferreUniversalMagneticDomain2013}. This would imply that ultrafast spin dynamics not only result in an overall demagnetization but can also affect the long-range spatial structure of magnetic domains over several tens of nanometers.

While ultrafast demagnetization is well established for a wide variety of ferromagnetic materials \cite{kirilyukLaserinducedMagnetizationDynamics2013, jeppsonCapturingUltrafastMagnetization2021}, only a few studies have hinted towards the ultrafast modification of nanoscale domain pattern \cite{pfauUltrafastOpticalDemagnetization2012, zusinUltrafastPerturbationMagnetic2022, zhouhagstromSymmetrydependentUltrafastManipulation2022}. These studies have used x-ray magnetic scattering to show that the diffraction rings obtained from labyrinthine domain pattern undergo ultrafast distortions of both ring radius and width. Tentative explanations have included domain wall broadening \cite{pfauUltrafastOpticalDemagnetization2012, santMeasurementsUltrafastSpinprofiles2017}, and the ultrafast rearrangement of domains \cite{zusinUltrafastPerturbationMagnetic2022, zhouhagstromSymmetrydependentUltrafastManipulation2022}. While these studies cannot clearly explain ultrafast distortions of diffraction patterns, domain rearrangement remains a viable hypothesis. 

\begin{figure*}[ht]
    \centering
    \includegraphics[width=\textwidth]{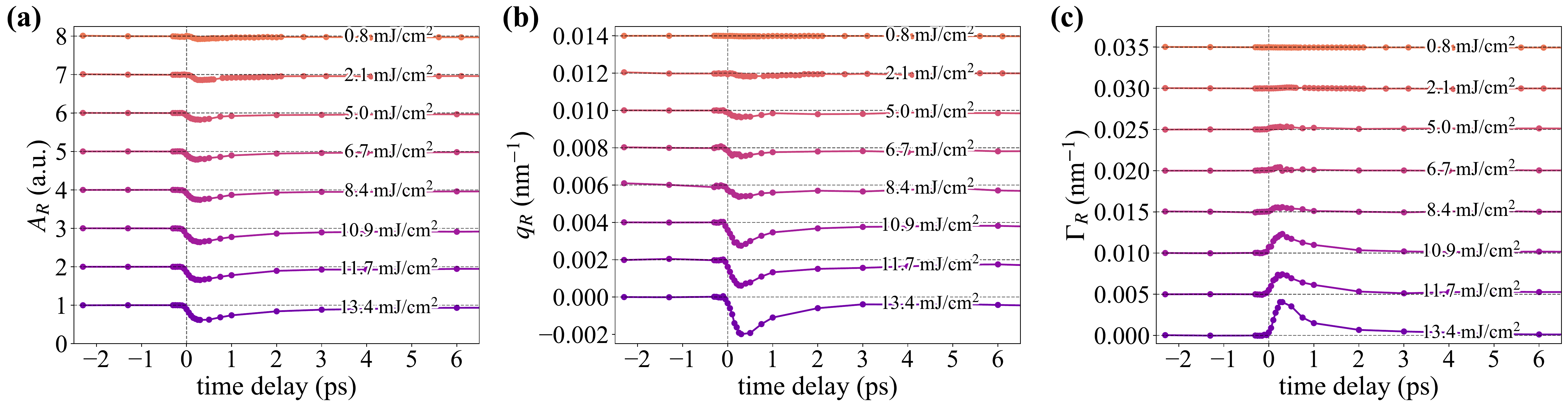}
    \caption{\textbf{Evolution of labyrinthine domain pattern as a function of delay time}. Time-resolved isotropic scattering including \textbf{(a)} amplitude ($A_R$), \textbf{(b)} ring radius ($q_R$) and \textbf{(c)} width ($\Gamma_R$) obtained from the 2D fit of the phenomenological model used for fitting the EUV diffraction pattern. Delay curves are plotted for a range of measured fluence values from 0.8 to 13.4 mJ/cm$^2$. The scattering amplitude which is proportional to magnetization, decays immediately following laser excitation indicating demagnetization which recovers on ps timescales. The ring radius ($q_R$) and width ($\Gamma_R$) of the isotropic scattering approximate the average real-space domain size and correlation length of the labyrinthine domains, respectively. Note that the plotted data for $A_R$, $q_R$ and $\Gamma_R$ is relative to the before t = 0 value.}
    \label{fig:IsoVStime}
\end{figure*}

To test the prediction of extreme-speed wall motion, we conducted optical pump, EUV (extreme ultraviolet) magnetic scattering probe experiments with a mixed-state domain pattern that consists of domains of both labyrinthine and stripe-like character. Scattering from such samples yields two dominant diffraction components; an azimuthally uniform and a twin-lobed ring pattern~\cite{zhouhagstromSymmetrydependentUltrafastManipulation2022}. We employed 2D fits similar to those in \cite{zhouhagstromSymmetrydependentUltrafastManipulation2022} to isolate and study the magnetization dynamics of domains of differing character. We measured the pump fluence dependence over an order of magnitude. Given that domain walls typically exhibit both inertia \cite{kittelNoteInertiaDamping1950, rhensiusImagingDomainWall2010} and an activation energy barrier, i.e. pinning \cite{rizzoRelaxationTimesMagnetization1999, ferreUniversalMagneticDomain2013}, the fluence dependence for the ultrafast distortion should be different from that of demagnetization if the ultrafast distortion is the result of domain-wall motion. We employed micromagnetic simulations to test the hypothesis that the preferential motion of curved domain walls in labyrinthine domains are in fact the source of ultrafast distortions. Our results provide experimental evidence for the theoretical proposition that far-from-equilibrium conditions can give rise to extreme domain-wall speeds.

Magnetic resonant scattering was measured by tuning the EUV photon energy to the M\textsubscript{3} edge of Ni at 66.2 eV at the FERMI free electron laser. Magnetic multilayered sample with stack layering of (Ta(3 nm)/Cu(5 nm)/[Co\textsubscript{90}Fe\textsubscript{10}(0.25 nm) /Ni(1.35 nm)] $\times$ 8 /Co\textsubscript{90}Fe\textsubscript{10}(0.25 nm)/Cu(5 nm)/Ta(3 nm)) were used. The sample was grown using magnetron sputtering on 100 nm thick polycrystalline Si membranes and is the same sample as the one used in \cite{zhouhagstromSymmetrydependentUltrafastManipulation2022}. Magnetic force microscopy (MFM) studies prior to the experiment showed the presence of linearly oriented labyrinth domains with an average width of 110 nm. 50 fs resolution pump-probe measurements were performed with an 800 nm pump and a linearly polarized EUV probe in transmission mode. The details about the experimental setup are included in SI section~\ref{Fitting_procedure}. Note that the nanoscale magnetic domain pattern of the sample (see figure~\ref{Sfig:MFM&setup}) exhibits two distinct diffraction features. The first is an isotropic ring attributed to the randomly twisting component of the labyrinth pattern. The second is an anisotropic lobes attributed to the stripe-like component of the domain pattern. In order to isolate the ultrafast behavior of these two features, the scattering data were fitted using a phenomenological 2D model similar to \cite{zhouhagstromSymmetrydependentUltrafastManipulation2022}. Additional details on the fitting procedure can be found in the SI section~\ref{Fitting_procedure}. 

Figure \ref{fig:IsoVStime} shows the ultrafast temporal evolution of the isotropic diffraction ring in terms of amplitude $A_R$, q-space radius $q_R$, and width $\Gamma_R$. An ultrafast distortion of the diffraction ring was observed, manifesting as both a reduction in the ring radius, and a broadening of the ring width. The temporal evolution of the equivalent parameters for the anisotropic lobe pattern ($A_L$, $q_L$, $\Gamma_L$) are presented in the SI figure~\ref{Sfig:AnisoVStime}. The demagnetization ($A_R$ and $A_L$) occurs within 100-200 fs followed by a slower recovery between 400 fs and 1.4 ps, depending on the fluence as further discussed below. A double-exponential fitting function as described in SI Section~\ref{Temporal_Fitting_procedure} was used to extract both magnitudes and time constants associated with the temporal response.

\begin{figure*}[ht]
    \centering
    \includegraphics[width = \textwidth]{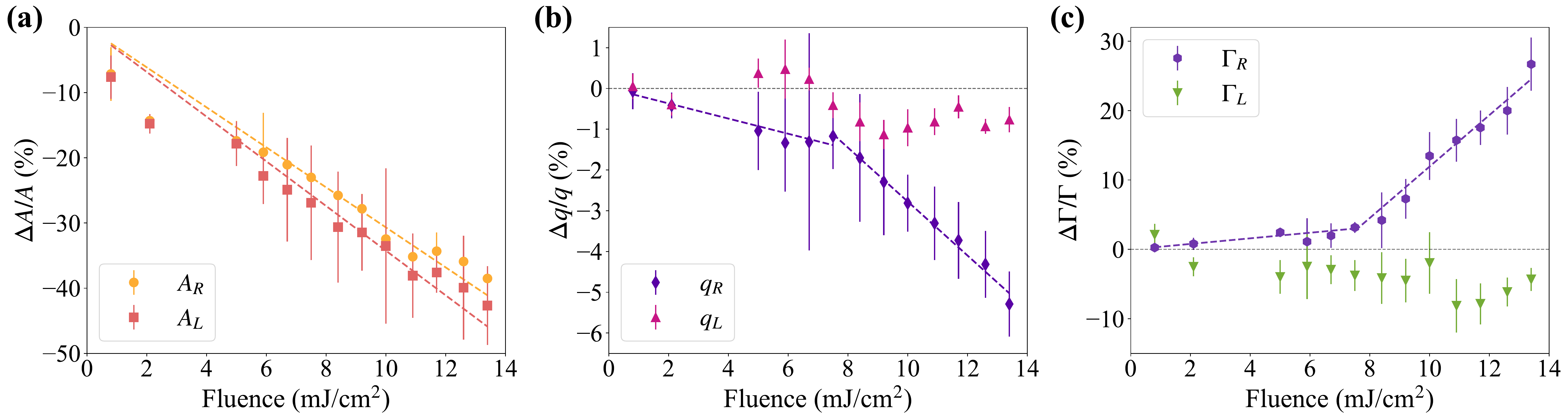}
    \caption{\textbf{Laser fluence dependence of isotropic and anisotropic scattering resulting from labyrinthine and stripe domains}. \textbf{(a)} Normalized scattered amplitude dependence on fluence for both the isotropic ($A_R$) and anisotropic scattering ($A_L$). Fluence depdence of \textbf{(b)} ring shift and \textbf{(c)} width for both the ring ($\Delta q_0/q_0$ and $\Delta \Gamma_0/\Gamma_0$) and lobes ($\Delta q_2/q_2$ and $\Delta \Gamma_2/\Gamma_2$). The dashed lines indicate the results of linear error-weighted fits of the data. For $A_R$ and $A_L$, the fits extend over the entire range of pump fluence. For $\Delta q_0/q_0$ and $\Delta \Gamma_0/\Gamma_0$, two fits were performed below and above the threshold fluence of 7.8 mJ/cm$^2$.}
    \label{fig:Aq_vs_fluence}
\end{figure*}

Figure \ref{fig:Aq_vs_fluence} shows the fluence-dependence for both the ring and lobes including demagnetization ($\Delta A_{R}/A_{R}$, $\Delta A_{L}/A_{L}$), radial peak shift ($\Delta q_R/q_R$, $\Delta q_L/q_L$) and ring broadening ($\Delta \Gamma_R/\Gamma_R$, $\Delta \Gamma_L/\Gamma_L$) relative to the average fitted pre-pump values for $t<0$. These results are also tabulated in Table \ref{Stab:Amp_table} in SI. The fluence dependencies of both $\Delta A_{R}/A_{R}$ and $\Delta A_{L}/A_{L}$ are very similar, and are consistent with most previous pump-probe studies \cite{zhouhagstromSymmetrydependentUltrafastManipulation2022}. The non-linearity of both $\Delta q_{R}/q_R$ and $\Delta \Gamma_{R}/\Gamma_R$ seen in Fig. \ref{fig:Aq_vs_fluence} (b) and (c) is in stark contrast to the linear fluence-dependence of the amplitude quenching $\Delta A_{R}/A_{R}$ and $\Delta A_{L}/A_{L}$ shown in Fig. \ref{fig:Aq_vs_fluence} (a). $\Delta q_{R}/q_R$ and $\Delta \Gamma_{R}/\Gamma_{R}$ exhibit a distinct threshold-like feature. For fluences below 7 mJ/cm$^2$, a relatively weak linear dependence of both $\Delta q_R/q_R$ and $\Delta \Gamma_R/\Gamma_R$ on fluence is observed. Above 7 mJ/cm$^2$, a much steeper linear dependence of $\Delta q_R/q_R$ and $\Delta \Gamma_R/\Gamma_R$ on pump fluence is observed, with $\Delta q_{R}/q_{R} = 5.3\pm0.8\%$ and $\Delta \Gamma_{R}/\Gamma_{R} = 26.7\pm3.8\%$ at the highest fluence. In contrast, $\Delta q_{L}/q_{L}$ and $\Delta \Gamma_{L}/\Gamma_{L}$ are much smaller and without any apparent linear dependence on fluence, with largest observed shifts of $1.0\pm0.4\%$ and $8.1\pm3.8\%$, respectively. The threshold-like behavior of the ultrafast diffraction ring distortions is the first main experimental result of this study. 

\begin{figure}[b!]
    \centering
    \includegraphics[width = 0.7\columnwidth]{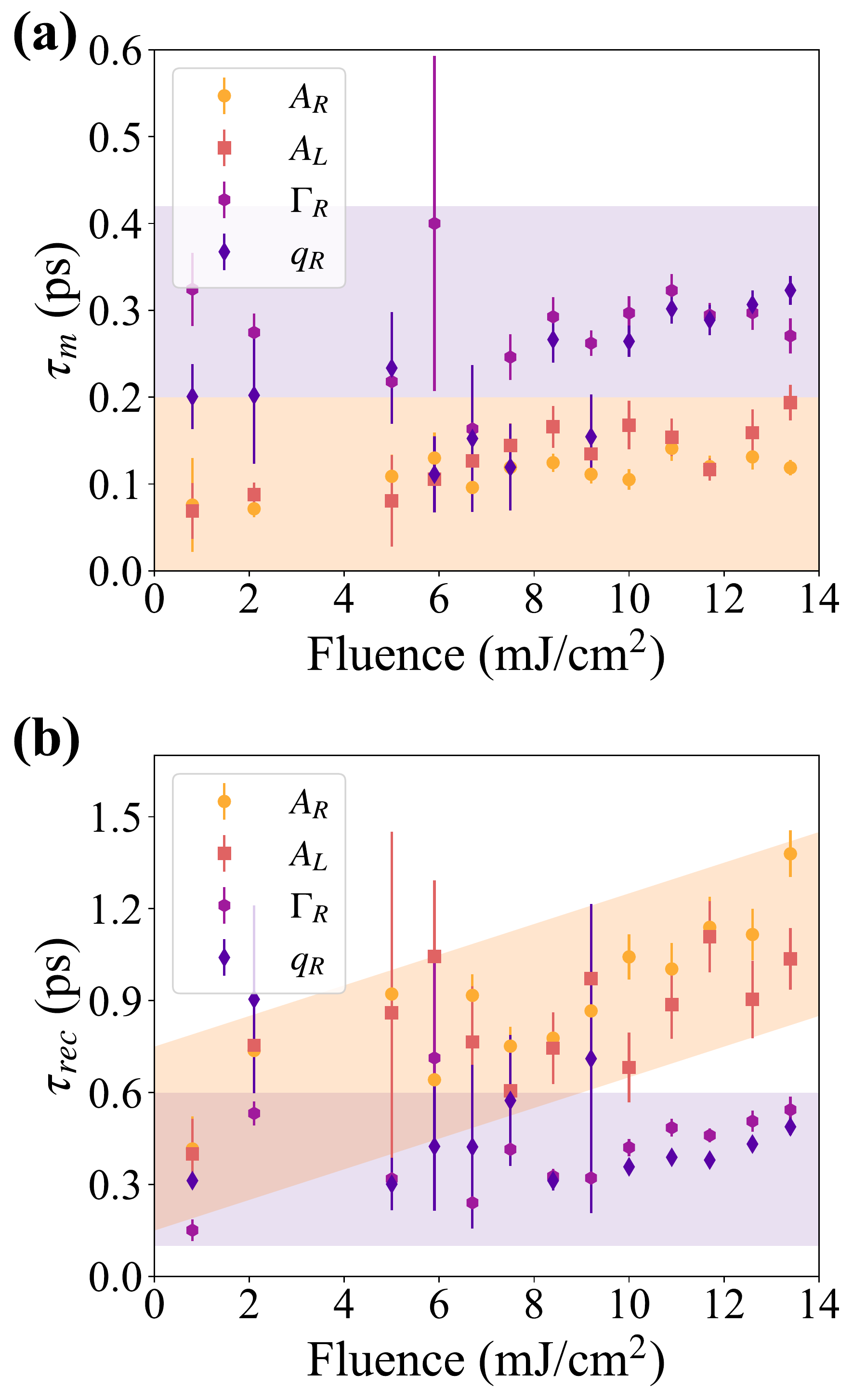}
    \caption{\textbf{Laser fluence dependence of quench and recovery time}. \textbf{(a)} Quench and \textbf{(b)} recovery time constants obtained from the temporal fits (see SI section \ref{Temporal_Fitting_procedure}) for  $A_R$, $A_L$, $q_R$ and $\Gamma_R$. The magnetization quench is two times faster than the change in radial ring position and ring width ($\tau_{m} \approx$ 0.3 ps) irrespective of the fluence value. The recovery time constants ($\tau_{rec}$) for magnetization quench ($A_R$ and $A_L$) are also distinct from $\tau_{rec}$ for ring shift ($q_R$) and width ($\Gamma_R$)}
\label{fig:t_min_vs_fluence}
\end{figure}

Time constants for the initial ultrafast changes $\tau_{m}$ and slower recovery $\tau_{rec}$ for $A_R$, $A_L$, $q_R$ and $\Gamma_R$ are presented in Figure \ref{fig:t_min_vs_fluence}. The demagnetization times for both the ring and lobes vary between 100 to 200 fs, indicative of a similar demagnetization process for labyrinths and stripes. Surprisingly, the time constants $\tau_{m}$ for the change in ring radius and ring width vary between 100 to 300 fs,  with most data falling between 200 and 300 fs; significantly slower than the demagnetization times. In addition, the recovery times $\tau_{rec}$ are also different between the demagnetization and ring shape distortions. The demagnetization recovery times vary from $\approx$ 600 fs to $\approx$ 1.2 ps, whereas both the ring radius and width recover much faster, with most data falling between 200 fs and 600 fs, dependent on the fluence. This difference in temporal response for demagnetization and ring distortion is the second key finding of this study. 

The threshold fluence for ring distortion ($\Delta q_{R}/q_{R}$ and $\Delta\Gamma_{R}/\Gamma_{R}$ in Figure \ref{fig:Aq_vs_fluence}) suggests that there is an activation energy barrier impeding domain rearrangement as typically observed for conventional field-driven wall dynamics \cite{ferreUniversalMagneticDomain2013}. This result is consistent with the hypothesis that domain rearrangement in the presence of pinning sites is the source of ultrafast ring distortions. Furthermore, the relatively slow rate (Fig. \ref{fig:t_min_vs_fluence}) for the change of $\Delta q_{R}/q_{R}$ and $\Delta \Gamma_{R}/\Gamma_{R}$ is consistent with domain-wall motion. Domain-walls are bound magnetic solitons that exhibit an effective inertia \cite{kittelNoteInertiaDamping1950, rhensiusImagingDomainWall2010} that impedes the response to any driving torque. Thus, based on the distinct response times for demagnetization and the ring distortions, we can confidently rule out any hypothesis that the distortions in diffraction ring shape are simply derivative results of ultrafast demagnetization process. 

Given the substantial differences in the ultrafast distortions of the stripe and labyrinth domain pattern, in agreement with the previous report \cite{zhouhagstromSymmetrydependentUltrafastManipulation2022}, it is natural to inquire what characteristic features of labyrinth and stripe domains underlie such differences in temporal response. An obvious difference is the abundance of curved domain walls for labyrinths. The possibility of curved wall motion, in contrast to that for straight walls, is consistent with the requirement of symmetry-breaking. Symmetry-breaking was provided in the original prediction of ultrafast wall motion by non-uniform laser illumination of a straight domain wall \cite{balazDomainWallDynamics2020}. In our case, the symmetry-breaking is geometrical, inherent in the wall curvature. 

\begin{figure}[b!]
    \centering
    \includegraphics[clip,width=0.95\columnwidth]{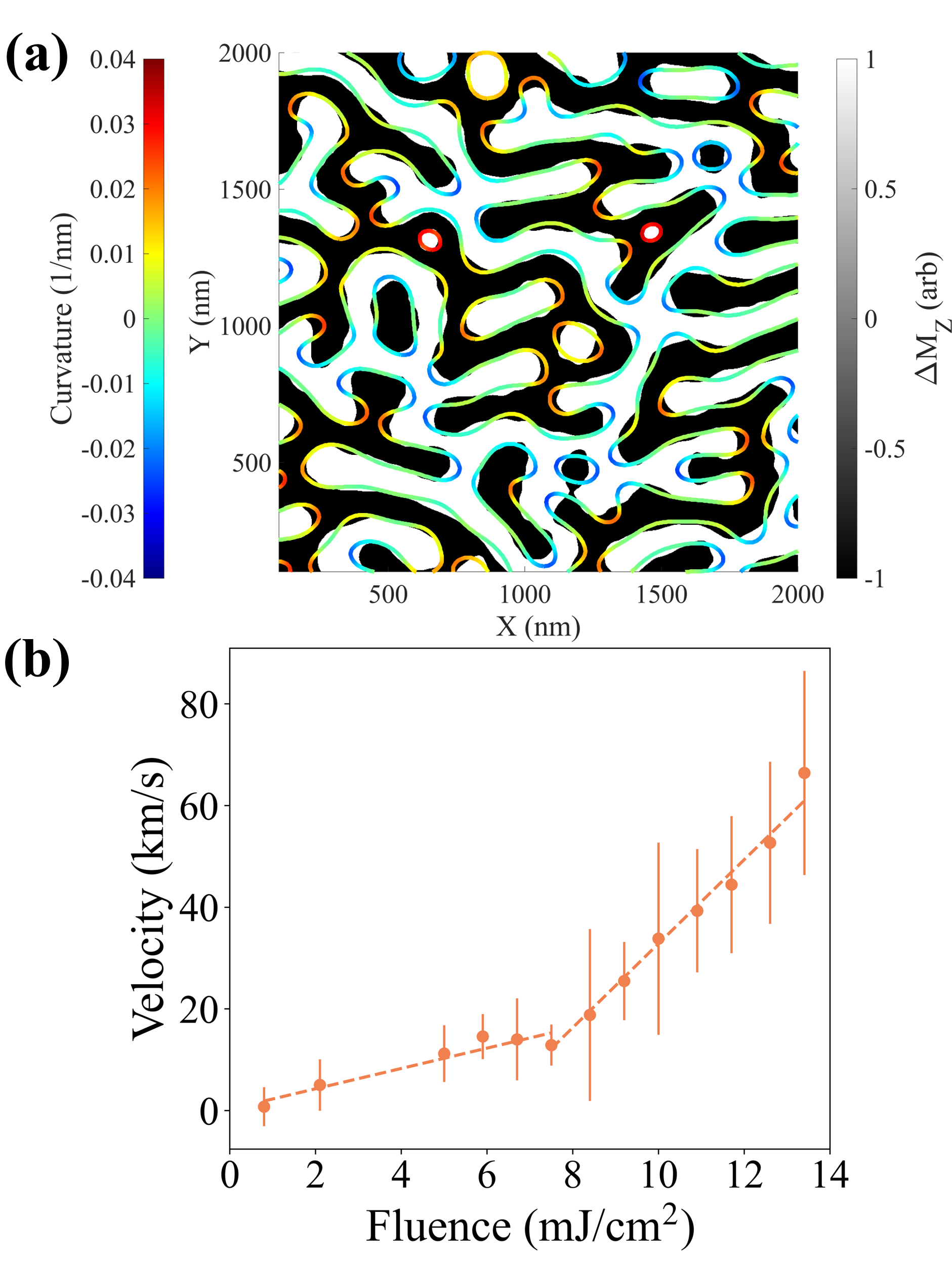}
    \caption{\textbf{Simulated modification of domain pattern and calculated domain wall velocity.} \textbf{(a)} Simulated modified domain pattern (black and white domains) and initial state (colored outline). The modified state was simulated assuming a 40 \% reduction in the saturation magnetization as discussed in the text. The color of the outline denotes the initial wall curvature which was estimated using inverse of the radius of local circle fit. The comparison clearly shows that regions with high curvature (dark red and blue) undergo noticeable domain wall motion. \textbf{(b)} Fluence dependence of calculated domain wall velocity for labyrinth domains estimated using experimentally measured and simulated contraction of diffraction ring radius.}
    \label{fig:Vel_vs_flu}
\end{figure}

To verify whether the spatial motion of curved domain walls can give rise to the observed contraction of the diffraction ring radius, we performed micromagnetic simulations \cite{vansteenkisteDesignVerificationMuMax32014} for the case of a perpendicularly magnetized thin film (see SI Section~\ref{Micromagnetic simulations}). An initial equilibrium domain state with either labyrinthine or stripe-like character was generated. The domain state was then modified by suddenly reducing the saturation magnetization by 40 \%. The magnetization was allowed to evolve according to conventional micromagnetic parameters to a new equilibrium state. This resulted in substantial domain wall displacements for the labyrinthine sample, where the displacements were proportional to the local domain wall curvature. While the time scale for the wall displacement is not accurate due to the use of micromagnetic simulations \cite{iacoccaSpincurrentmediatedRapidMagnon2019},  the dependence of wall displacement on the curvature allows us to examine how such domain rearrangement affects the domain pattern in reciprocal space. Figure \ref{fig:Vel_vs_flu}(a) presents the simulated modified labyrinth domain pattern (black and white domains) and compares it with the initial domain pattern, where only the outline is shown and the color denotes initial wall curvature. The figure clearly shows that domain walls with higher curvature (dark red or blue) undergo noticeable wall motion. Both the observed diffraction ring radius contraction and width broadening were qualitatively reproduced by FFT analysis of the modified simulated labyrinth domain state (see SI Section~\ref{Micromagnetic simulations}). In contrast, the same modeling of a stripe domain pattern with minimal curved walls does not show any detectable distortions in the shape of the FFT spatial pattern. 

We can now estimate the domain wall speeds of the curved walls by utilizing both the experimentally measured and simulated contraction of diffraction ring radius. This was achieved by quantitatively correlating the reduction of diffraction ring radius with the modeled change in the wall position. We then determined the average wall displacement necessary to cause the experimentally observed contraction in the diffraction ring radius (see SI section~\ref{Velocity Calculation}). Combined with the experimentally measured radial contraction times, we can then estimate average domain wall speeds of the curved walls as a function of pump fluence, presented in figure \ref{fig:Vel_vs_flu}(b). For 13.4 mJ/cm$^2$, we estimated rms curve wall displacement to be $\approx$ 20 $\pm$ 3 nm, which results in the calculated domain wall speed of 66 $\pm$ 20 km/s. It was recently shown that wall speeds approaching the maximum magnon group velocity are physically allowed for a ferrimagnet under equilibrium conditions \cite{carettaRelativisticKinematicsMagnetic2020}. In our case, the maximum group velocity for Ni is $\approx$ 63 km/s \cite{mookNeutronScatteringMeasurementSpinWave1985}. Thus, we also find that wall speeds approaching the maximum magnon velocity are also possible for curved domain walls in a ferromagnet, but under extreme far-from-equilibrium conditions. It should be also noted that the wall motion could be a combination of both wall motion and broadening, and our observations do not rule out domain wall broadening previously observed \cite{pfauUltrafastOpticalDemagnetization2012, zusinUltrafastPerturbationMagnetic2022}. The observation of extreme wall speed under far-from-equilibrium conditions is the third and most significant result of this study. 

We note that the observation of threshold effect and distinct time constants of ring distortion and demagnetization indicate that the existing theory is still inadequate to predict the scale of the observed phenomena. The faster rate of the distortion recovery suggests more complex physics whereby the relaxation channels for wall dynamics are not identical to those for the demagnetization recovery. It is possible that other mechanisms such as magnon excitation and relaxation need to be included \cite{zhangRelaxationTimeTerahertz2012, zakeriElementarySpinExcitations2014, elhanotyElementselectiveUltrafastMagnetization2022}. It has already been shown that ultrafast demagnetization results in substantial magnon generation~\cite{choiSpinCurrentGenerated2014, turgutStonerHeisenbergUltrafast2016, knutInhomogeneousMagnonScattering2018, beensModelLocalNonlocal2020}. Enhanced demagnetization in domain walls has been attributed to the excitation of both coherent and incoherent magnon-like modes in chiral domain walls \cite{leveilleUltrafastTimeevolutionChiral2022}. Indeed, the observed temporal response of the wall dynamics are similar to that of critically damped oscillator. This suggests that far-from-equilibrium conditions can give rise to new sources of elastic torque that can affect mesoscopic spin textures in a coherent manner. 

Our work highlights two critical points for far-from-equilibrium behavior. First, our results show significant evidence of extreme domain walls speeds in qualitative agreement with theoretical predictions \cite{balazDomainWallDynamics2020}. Furthermore, we show that the conventional formulation of magnetic torques is inadequate to account for all experimental observations, especially recovery timescales. The usual theory for domain walls in ferromagnets is micromagnetic, where torques arise from energy terms in the GHz range when constrained to mesoscopic scales (10 to 100 nm). For extreme wall motion to occur, micromagnetic energy terms on the order of meV activated under far-from-equilibrium conditions are required. This is a surprising result since domain walls in ferromagnets near equilibrium are unstable when driven above the Walker limit \cite{schryerMotion180Domain1974, ferreUniversalMagneticDomain2013}. Dissipative superdiffusive spin current is a possible source of the requisite THz-scale energy, but our distortion-recovery data show that it cannot be the only relevant mechanism; meV-scale elastic terms are required. Secondly, most proposed mechanisms for ultrafast demagnetization rely on entropy-producing microscopic single-particle processes. Such processes occur on a length scale between the lattice constant and the exchange length. However, extremely fast spatial translation of domain walls requires a long-range mechanism that extends over tens of nanometers, i.e at the mesoscopic scale \cite{heydermanMesoscopicMagneticSystems2021}. The implication is that far-from-equilibrium spin kinetics in ferromagnets are not solely limited to demagnetization mechanism. There must also be generation of coherent torques at ultrafast time-scales in non-uniform mesoscopic spin textures. Our studies thus open up the possibility of manipulating magnetic textures to achieve far-from-equilibrium mesoscopic effects. Furthermore, the extension of these processes could be important for explaining ultrafast phenomena in other systems such as emerging quantum materials.

\section*{Acknowledgments}
The authors acknowledge the FERMI Free Electron laser in Trieste, Italy for allowing us to use the Diffraction and Projection Imaging (DiProI) beamline and thank the beamline scientists and facility staff for their assistance. R.J., M.M. and R.K. acknowledge support from AFOSR Grant. No. FA9550-19-1-0019. N.Z.H. and S.B. acknowledge support from the European Research Council, Starting Grant 715452 MAGNETIC-SPEED-LIMIT. E.I. acknowledges the College of Letters, Arts and Sciences at UCCS for start-up support. This material is based upon the work supported by the National Science Foundation under Grant No. 2205796 and this research was funded by National Institute of Standards and Technology \href{https://ror.org/05xpvk416}{(NIST)}. The authors acknowledge D. Bozhko for representative experimental data to seed stripe domains in micromagnetic simulations. Certain equipment, instruments, software, or materials, commercial or non-commercial, are identified in this paper in order to specify the experimental procedure adequately. Such identification is not intended to imply recommendation or endorsement of any product or service by NIST, nor is it intended to imply that the materials or equipment identified are necessarily the best available for the purpose. The raw data generated at the FERMI and the code are available from the corresponding author upon reasonable request.

\section*{Supplementary Information}
\beginsupplement

\subsection{\label{Fitting_procedure} 2D fitting procedure and results}
\begin{figure}[ht!]
    \includegraphics[width = 0.45\textwidth]{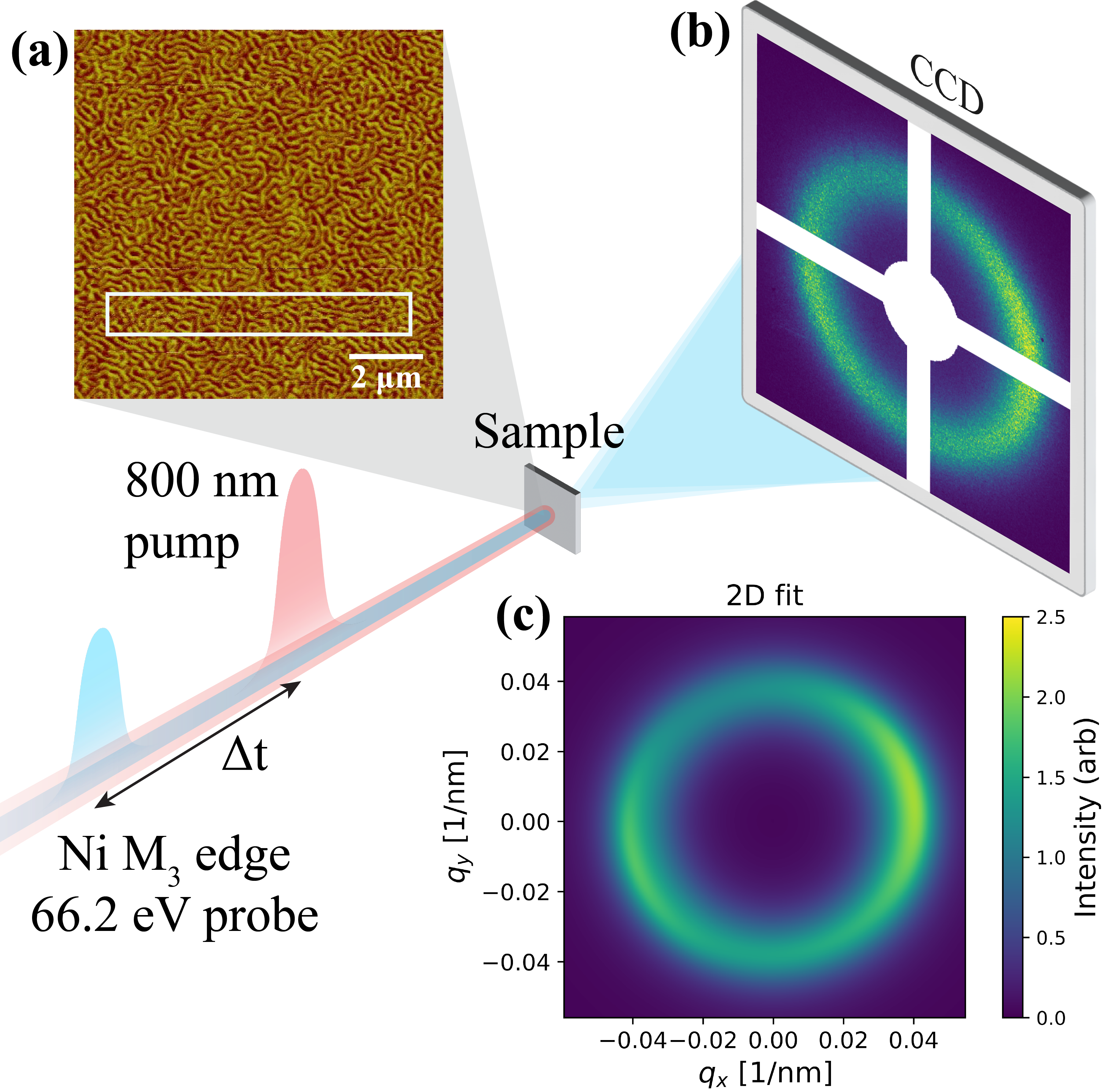}
    \caption{\textbf{Schematic of the optical pump EUV magnetic scattering probe setup}. Time-resolved studies were performed using 800 nm pump and 66.2 eV (Ni M\textsubscript{3} edge) probe. \textbf{(a)} MFM image (10 \textmu m $\times$ 10 \textmu m FOV) of the sample. The white box highlights the horizontal linear texture of the domain pattern. \textbf{(b)} Magnetic diffraction pattern from the sample on the CCD. \textbf{(c)} 2D fit results as described in the text for the scattering data shown on the CCD.}
    \label{Sfig:MFM&setup}
\end{figure}

The schematic of time-resolved magnetic scattering setup is shown in \ref{Sfig:MFM&setup} along with the MFM image of the magnetic domains measured prior to the FERMI experiment.  As mentioned in the main article, the diffraction pattern consisted of both isotropic and anisotropic scattering components due to the presence of both labyrinthine and stripe domains. The laser pump fluence was calculated assuming a flat top profile for the energy density and a spot size of 390 \textmu m. EUV fluence was limited to 1 mJ/cm$^2$ to prevent any pumping effects from it. In order to extract these components from the diffraction we utilized 2D fitting using a phenomenological function as described below. 

\begin{figure}[ht]
    \centering
    \includegraphics[width = 0.8\columnwidth]{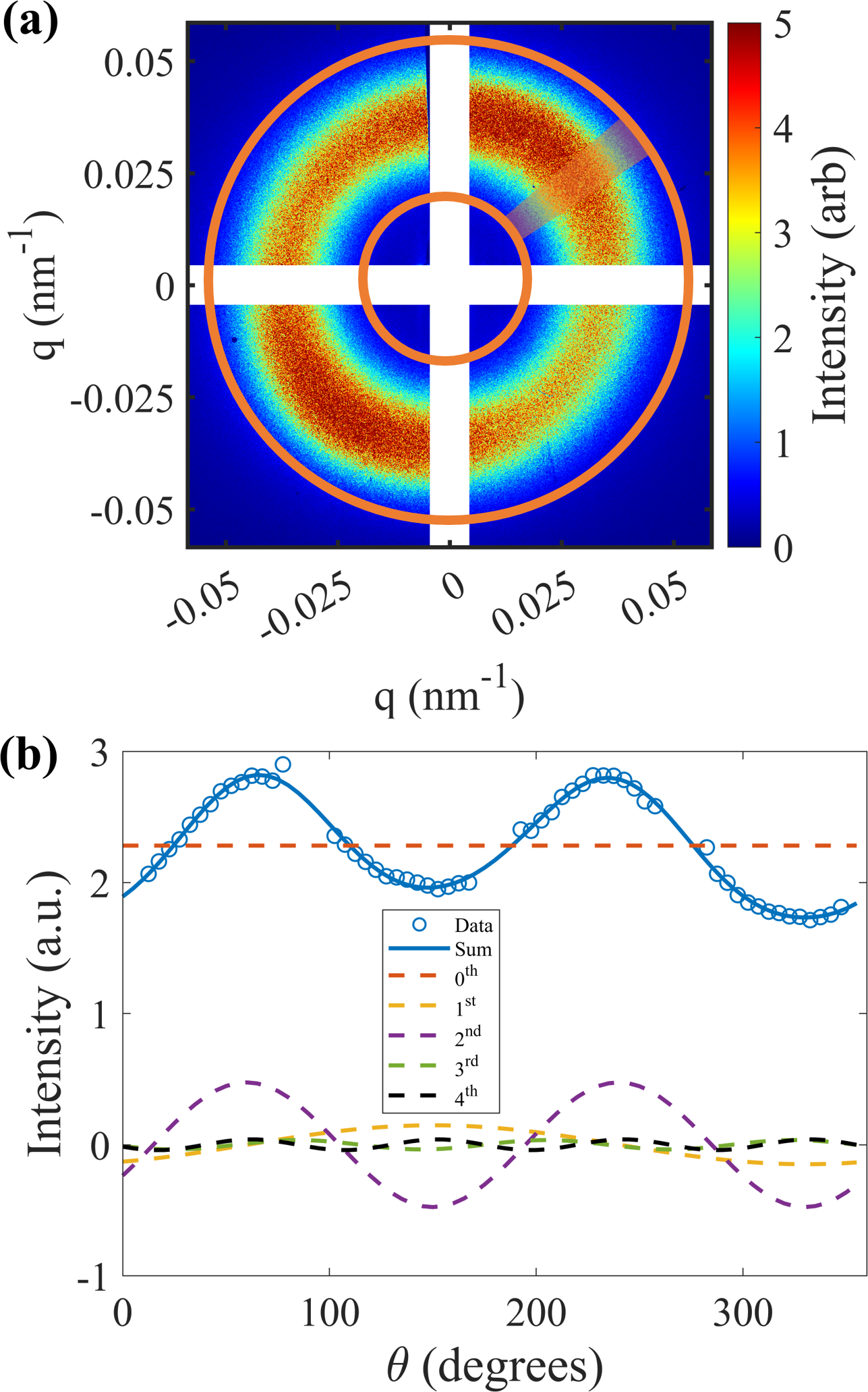}
    \caption{\textbf{Intensity as a function of azimuthal angle and its Fourier decomposition}. \textbf{(a)} Representative scattering pattern from a mixed state along with a schematic of the wedge used to create an azimuthal profile. \textbf{(b)} Azimuthal profile (blue data points) and fit (blue line) along with the individual Fourier series components ($0^{th}$ to $4^{th}$).}
    \label{Sfig:IvsAzimuth_angle}
\end{figure}

In order to establish the phenomenological function used for the 2D fitting, we employed Fast Fourier Transforms (FFTs) of the radially averaged diffraction intensity along the azimuthal dimension. Figure \ref{Sfig:IvsAzimuth_angle}(a) shows a scattering image (post background subtraction) along with a schematic of a wedge over which the data was averaged to extract the azimuthal profile. The obtained azimuthal profile is plotted in figure \ref{Sfig:IvsAzimuth_angle}(b) along with the Fourier series fit components. The FFT obtained from the measured diffraction pattern exhibited five distinct peaks, identified as the 0$^{th}$ through 4$^{th}$ azimuthal harmonics. 0$^{th}$ (isotropic scattering), 1$^{st}$ (odd harmonic scattering), and 2$^{nd}$ (anisotropic scattering) order were found to be the major components in the azimuthal dimension. The observed 0$^{th}$ and 2$^{nd}$ harmonics indicate that the sample is in a mixed state with both labyrinthine and stripe-like components to the domain pattern (see figure \ref{Sfig:MFM&setup}(a)). This is consistent with the recent results reported for the same samples measured at the L$_3$ edge at the European XFEL by \citet{zhouhagstromSymmetrydependentUltrafastManipulation2022}. The observed 1$^{st}$ harmonic may originate from birefringence due to the chiral nature of the Bloch-like domain walls. Its origin is discussed in detail in the paragraph below. The observed 3$^{rd}$ and 4$^{th}$ harmonics are higher-order components of the azimuthal dependence of the diffraction pattern with an order of magnitude lower amplitude compared to the 0$^{th}$ and 2$^{nd}$ orders.

Based on the FFT analysis, five components with each proportional to $\cos{(n\theta)}$, where $n=0...4$ were included in the phenomenological model used for fitting the scattered diffraction pattern. A symmetric Lorentzian in the radial direction modulated with a Fourier cosine term in the azimuthal dimension was found to give the best fits with the lowest residual. The functional form used to fit the 2D scattering images is given by,  
\begin{equation}
    f(q, \varphi) = B + f_{iso}(q) + f_{odd}(q, \varphi) + f_{aniso}(q, \varphi)
    \label{eqn1:2Dfit_func}
\end{equation}
where $B$ is the uniform background, $q$ is the wavevector, and $\varphi$ is the azimuthal angle. $f_{iso}(q)$ represents the isotropic component (0$^{th}$ harmonic) of the scattering which is a result of scattering from the randomly oriented labyrinthine domains. It is a function of wavevector $q$ and is modeled using a symmetric Lorentzian (see Eqn \ref{eqn2:f_iso}),

\begin{equation}
    f_{iso}(q) = I_{R}\left[\frac{1}{\left[\frac{\left(q-q_{R}\right)}{\Gamma_{R}}\right]^{2}+1}\right]^{2}.
    \label{eqn2:f_iso}
\end{equation}

where $q_R$ is the ring radius and $\Gamma_R$ is the ring width. $f_{odd}(q, \varphi)$ is the asymmetric scattering component (1$^{st}$ and 3$^{rd}$ harmonics) and is a function of both wavevector ($q$) and the azimuthal angle ($\varphi$). $f_{odd}(q, \varphi)$ is modeled as a symmetric Lorentzian modulated by the appropriate odd-order harmonics (see equation \ref{eqn3:f_odd}).

\begin{equation}
    \begin{aligned}
        & f_{odd}(q, \varphi) = \left[\frac{1}{\left[\frac{\left(q-q_{O}\right)}{\Gamma_{O}}\right]^{2}+1}\right]^{2}\times \\
        & \left[\left[\frac{I_{O}}{2}\left(\cos \left(\theta-\varphi_{O}\right)+1\right)\right] + \left[\frac{I_{O_3}}{2}\left(\cos \left(3\left(\theta-\varphi_{O}\right)\right)+1\right)\right]\right].
    \end{aligned}
    \label{eqn3:f_odd}
\end{equation}

The anisotropic scattering which originates from the stripe-like domains (2$^{nd}$ and 4$^{th}$ harmonic) is represented by $f_{aniso}(q, \varphi)$. It is also a function of wavevector $q$ and the azimuthal angle $\varphi$ and is modeled using a symmetric Lorentzian modulated with the even-order cosine harmonics:

\begin{equation}
    \begin{aligned}
        & f_{aniso}(q, \varphi) = \left[\frac{1}{\left[\frac{\left(q-q_{L}\right)}{\Gamma_{L}}\right]^{2}+1}\right]^{2}\times \\
        & \left[\left[\frac{I_{L}}{2}\left(\cos \left(2\left(\theta-\varphi_{L}\right)\right)+1\right)\right]+\left[\frac{I_{L_4}}{2}\left(\cos \left(4\left(\theta-\varphi_{L}\right)\right)+1\right)\right]\right].
    \end{aligned}
    \label{eqn4:f_aniso}
\end{equation}

\begin{figure*}[ht!]
    \centering 
    \includegraphics[width=\textwidth]{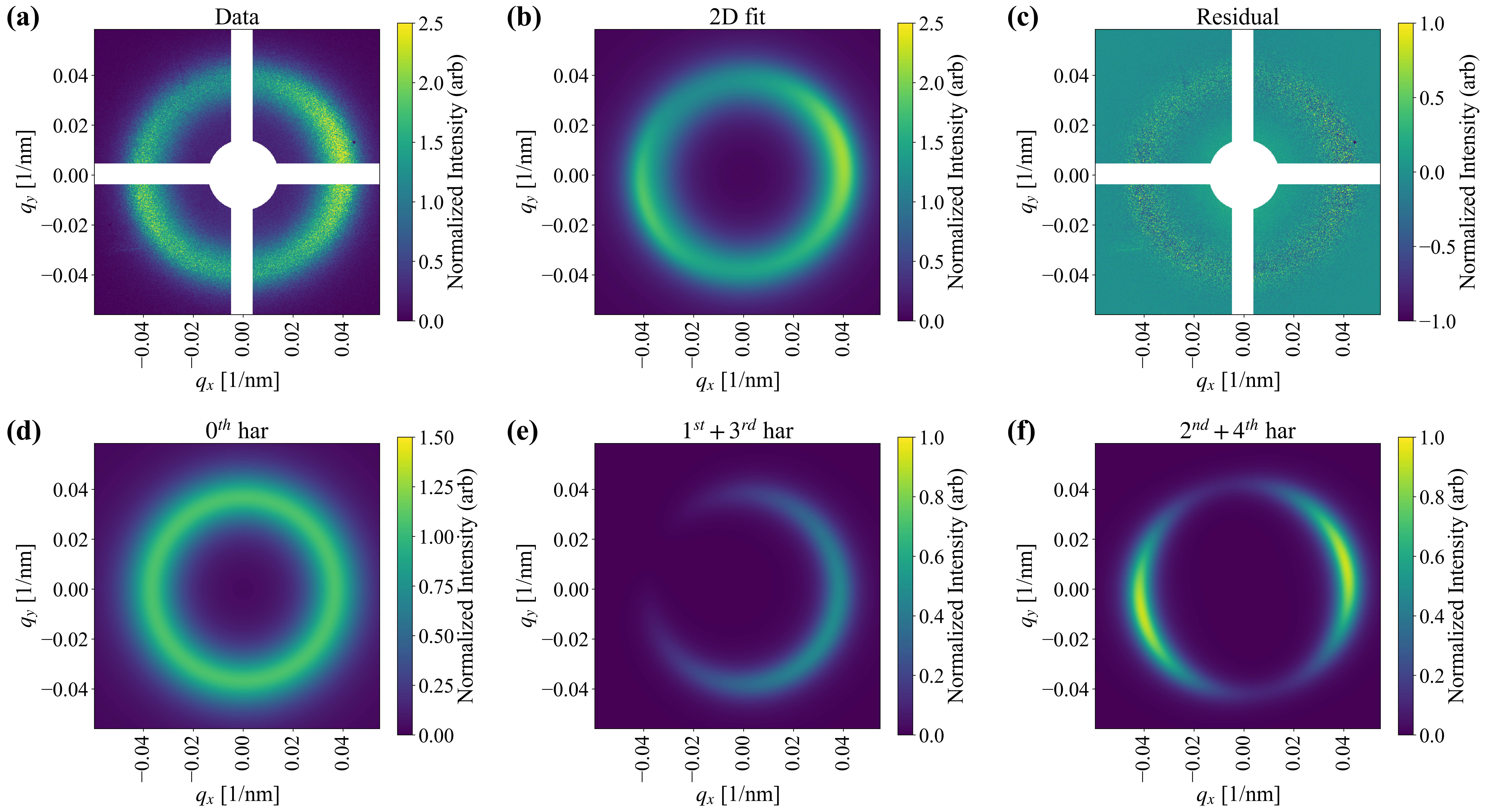}
    \caption{\textbf{2D fit results and fit components}. Shows 2D fit results for a representative scattering pattern using the phenomenological model. \textbf{(a)} raw  experimental scattering data, \textbf{(b)} fit results, and \textbf{(c)} residual. Note that the intensity scale for residual is 2.5 times smaller than the raw data and the fit. Isolated scattering obtained from the fits for \textbf{(d)} labyrinth domains, \textbf{(e)} odd harmonic, and \textbf{(f)} stripe domains. Note that the intensity scale for odd harmonic and stripe domains is 1.5 $\times$ smaller than the scale for labyrinth domains.}
    \label{Sfig:2D_fit_results_and_components}
\end{figure*}

The final form of the 2D fit equation was obtained by combining Eqns \ref{eqn1:2Dfit_func}, \ref{eqn2:f_iso}, \ref{eqn3:f_odd} and \ref{eqn4:f_aniso}. Figure \ref{Sfig:2D_fit_results_and_components} shows the results from the aforementioned 2D fit procedure along with the residual. Figure \ref{Sfig:2D_fit_results_and_components}(d-f) shows the individual components from the 2D fit. Scattering from the labyrinthine domains has the highest amplitude with twice the intensity and smaller radius ($q_R$) compared to the other components. This difference in the wavevectors for labyrinth domains compared to other components gives rise to an elliptical appearance of the scattering pattern in Figure \ref{Sfig:2D_fit_results_and_components}(a). The 2$^{nd}$ order was the second largest component, with roughly half the intensity of the 0$^{th}$ order component. The functional form of our 2D fit function is similar to the phenomenological model used by \citet{zhouhagstromSymmetrydependentUltrafastManipulation2022}. The primary difference is due to FFT being applied to obtain the azimuthal dependence of the diffraction pattern intensity. Thus, we fitted intensity and not amplitude. As such, all reported values for amplitude $A$ are the magnitude $|A|$, derived from the intensity $I$ via $A=\sqrt{I}$. Note that the intensity scale for the residuals is 2.5 times smaller than the raw data and the fit. The small amplitude of the residual indicates that our model is an adequate approximation of the data for the purposes of time-resolved analysis

\begin{figure*}[ht]
    \centering
    \includegraphics[width = \textwidth]{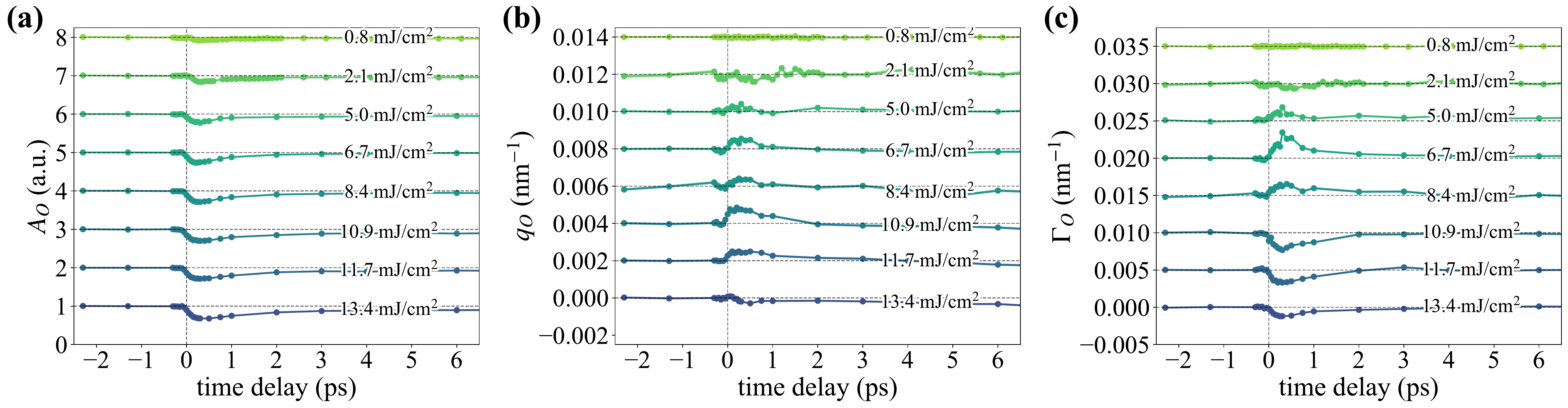}
    \caption{\textbf{Evolution of odd harmonic scattering as a function of delay time and pump fluence}. \textbf{(a)} scattering amplitude $A_O$, \textbf{(b)} peak position $q_O$, and \textbf{(c)} peak width $\Gamma_O$ for the $1^{st}$ harmonic as a function of time delay for various pump fluences obtained from 2D fit of the phenomenological model used for fitting the diffraction pattern.}
    \label{Sfig:OddVStime}
\end{figure*}

\begin{figure*}[ht]
    \centering
    \includegraphics[width = \textwidth]{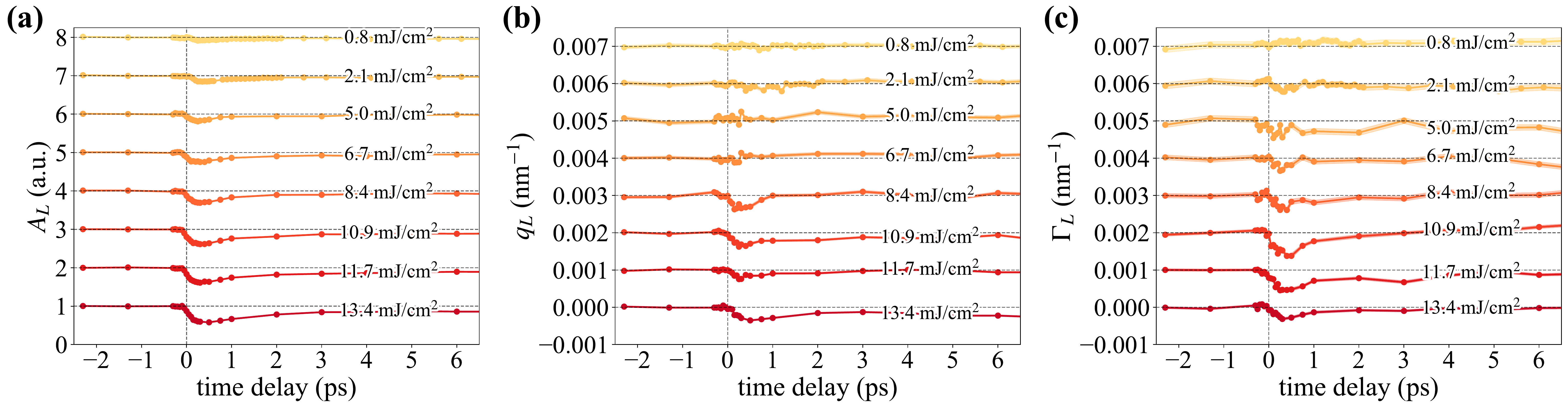}
    \caption{\textbf{Evolution of anisotropic scattering from stripe domains as a function of delay time}. \textbf{(a)} Scattering amplitude $A_L$, \textbf{(b)} peak position $q_L$, and \textbf{(c)} peak width $\Gamma_L$ for $2^{nd}$ harmonic as a function of time delay for various pump fluences obtained from 2D fit of the phenomenological model used for fitting the diffraction pattern.}
    \label{Sfig:AnisoVStime}
\end{figure*}

\begin{figure}[b]
    \centering
    \includegraphics[width = 0.7\columnwidth]{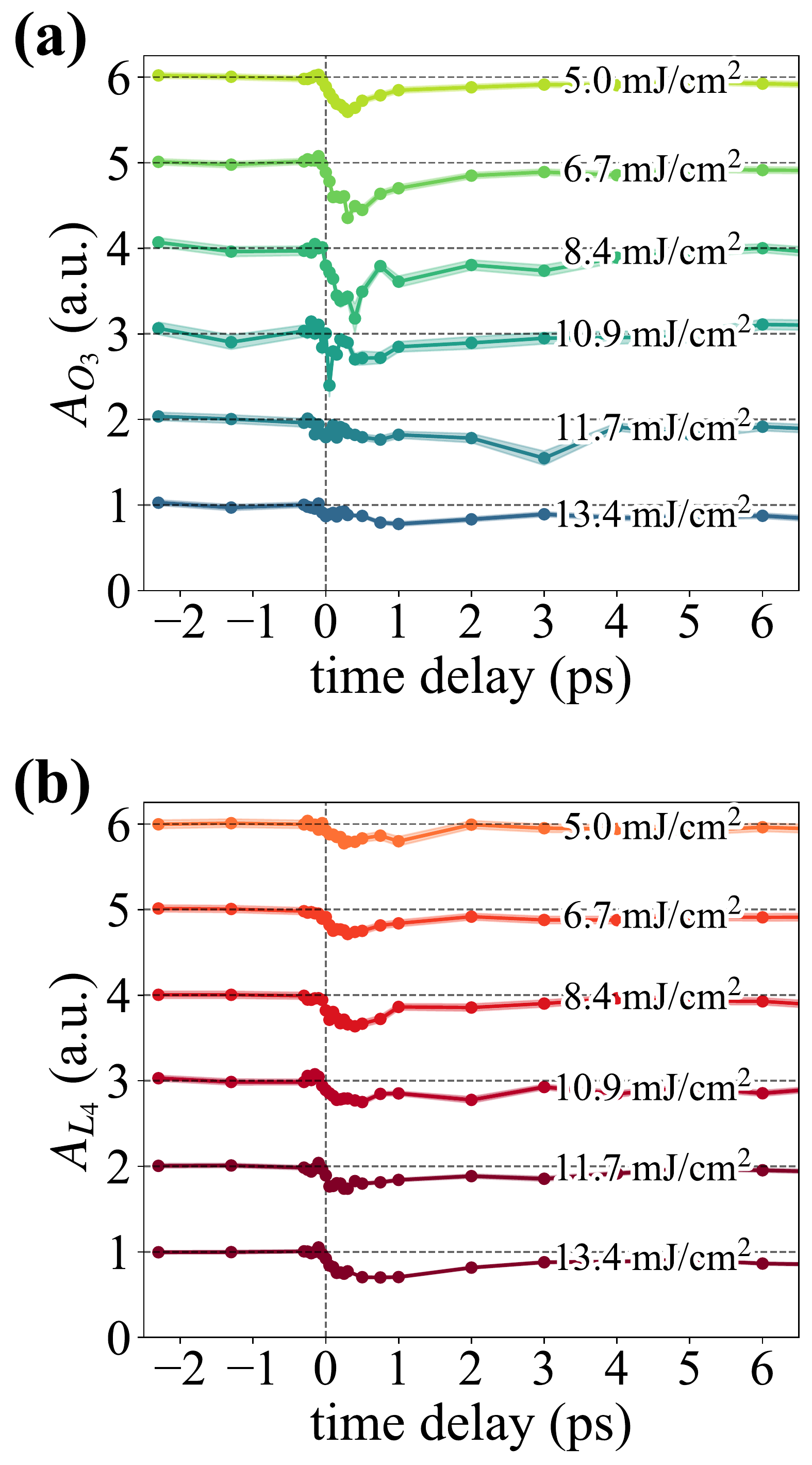}
    \caption{\textbf{Scattering amplitude of \boldsymbol{$3^{rd}$} and \boldsymbol{$4^{th}$} harmonics as a function of delay time for various pump fluences.}}
    \label{Sfig:A3_A4_vs_dt}
\end{figure}

Figure \ref{Sfig:OddVStime} and \ref{Sfig:AnisoVStime} show the time-dependence of amplitudes ($A_{O}, A_{L}$), peak position ($q_O, q_L$) and peak width ($\Gamma_O, \Gamma_L$) for the odd and even azimuthal harmonics, respectively, obtained from the 2D fits. The amplitudes of magnetization quench for both $A_O$ and $A_L$ show similar laser fluence dependence as $A_R$ (Figure \ref{fig:IsoVStime}(a)). The ultrafast responses of $q_O$, $q_L$, $\Gamma_O$ and $\Gamma_L$ are relatively small compared to the $0^{th}$ order components (Figure \ref{fig:IsoVStime}(b-c)). The amplitudes of the $3^{rd}$ and $4^{th}$ harmonics are 10 times smaller than the $1^{st}$ and $2^{nd}$ harmonics as shown in Fig \ref{Sfig:A3_A4_vs_dt}. Note that low fluences are not shown in Fig \ref{Sfig:A3_A4_vs_dt} due to inadequate signal-to-noise. 

As mentioned above, the FFT analysis also yielded two \emph{odd-order} components. From the standpoint of diffraction alone, such odd-order diffraction features are forbidden. One plausible explanation for these anomalous components is the combination of the transverse magneto-optic effect in transmission combined with far-field diffraction from domain walls. Under the assumption that the walls are Bloch-like, an in-plane component of magnetization in the domain wall extends through much of the sample thickness. When the linear polarization of the light is orthogonal to the in-plane component of the wall magnetization, the in-plane magnetization gives rise to asymmetric scattering via magnetic birefringence in the EUV regime. The physical mechanism of such an effect is analogous to the transverse magneto-optic Kerr effect (T-MOKE), where the reflectivity of p-polarized light is proportional to magnetization when $x-z$ plane of incidence is perpendicular to the magnetization along the $y$ axis. In the case of T-MOKE, birefringence rotates the surface normal component of the optically-induced polarization $P_z$ into the $x$-direction, so that it either adds to, or subtracts from, the specular far-field reflection, depending on the strength of the effect and the direction of the magnetization. In the case of transmission for $x$-polarized light with uniform magnetization along the $y$-axis, birefringence results in a surface-normal component of the polarization $P_z$. However, far-field transmission as a result of birefringence alone is forbidden because the in-plane component of the optical wavevector is zero, and, therefore, $P_z$ is parallel to the wavevector of the transmitted light. This is why there isn't an analogue of T-MOKE in transmission. However, if the magnetization is non-uniform such that the wavevector of the transmitted light now has an in-plane component, contribution of the surface-normal $P_z$ to diffraction is now allowed. This effect is a magnetic analogue to that of a blazed diffraction grating, where diffraction and reflection are collinear. A similar one-sided lobe structure via diffraction from a perpendicular multi-domain sample was previously observed by \citet{santMeasurementsUltrafastSpinprofiles2017} in the reflection geometry. 

\begin{figure*}[ht]
    \centering
    \includegraphics[width = 0.7\textwidth]{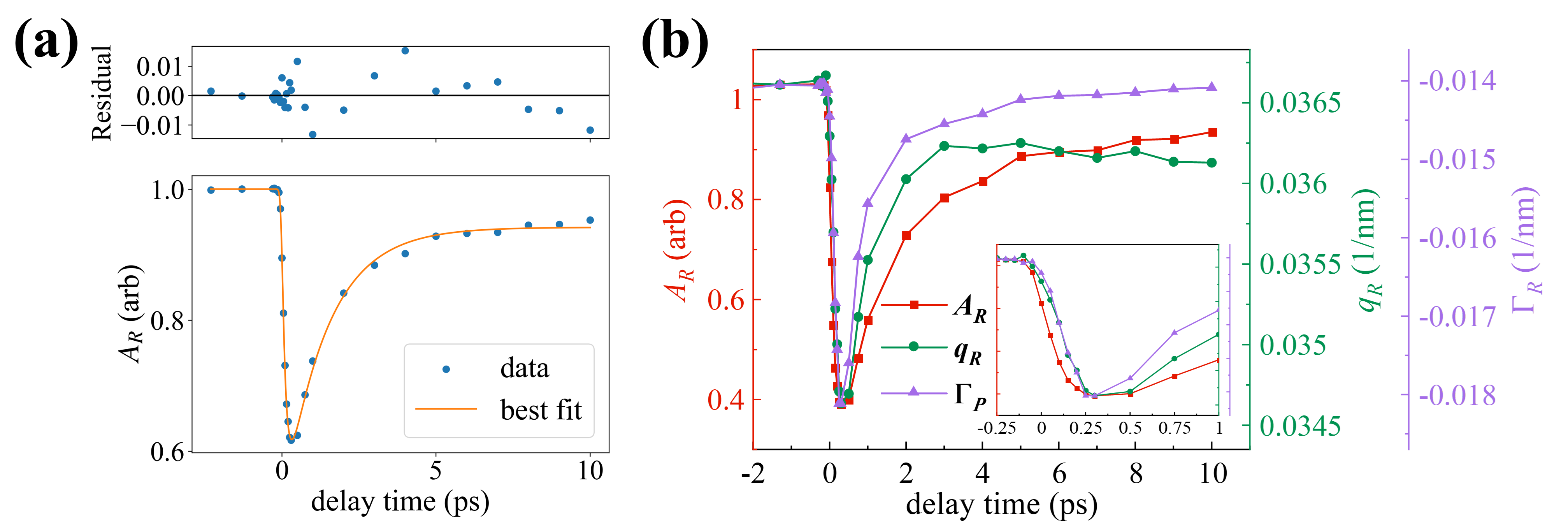}
    \caption{\textbf{Temporal fit and comparison of temporal evolution of \boldsymbol{$A_R$}, \boldsymbol{$q_R$} and \boldsymbol{$\Gamma_R$}}. \textbf{(a)} Temporal evolution of scattering amplitude for labyrinth domains along with the fit for 13.4 mJ/cm$^2$. This fit was obtained using a temporal fit function discussed in SI section \ref{Temporal_Fitting_procedure}. The residual from the fit is also shown above. \textbf{(b)} Compares the temporal evolution of the scattering amplitude ($A_R$) radial ring position ($q_R$) and the ring width ($\Gamma_R$) for labyrinth domains for a fluence of 13.4 mJ/cm$^2$.}
    \label{Sfig:timescales_and_fit}
\end{figure*}

In our case, the observation of such an effect is contingent on a trivial magnetic topology, where the in-plane component of the magnetization does not alternate periodically between adjacent walls, i.e. adjacent walls have opposite chirality. Otherwise, the sense of gyromagnetic rotation of the optical polarization in the walls would alternate sign, thereby destructively interfering when transmitted into the far field. Given the nature of our samples, the requisite Dzyaloshinskii-Moriya interaction (DMI) necessary to imbue non-trivial topology is not expected.
Odd-order components were not observed when the same samples were utilized at a coherent X-ray source \cite{zhouhagstromSymmetrydependentUltrafastManipulation2022}. This is to be expected because of the greatly reduced Faraday rotation of magnetic materials at x-ray wavelengths \cite{valenciaFaradayRotationSpectra2006, kortrightSoftxrayFaradayRotation1995}. Further elucidation of odd-order is required to verify the physical origins of the effect. Our FFT analysis also indicates the presence of a 3$^{rd}$ and 4$^{th}$ order harmonics. However, these higher-order components were $>$10 times smaller than the 1$^{st}$ and 2$^{nd}$ order harmonics. Subsequent analysis shows that they react very similarly when optically pumped. Given their small amplitudes, as well as the apparent redundancy of their ultrafast response, we focused on the 0$^{th}$ and 2$^{nd}$ harmonics. 

\subsection{\label{Temporal_Fitting_procedure}Time constant fits and results}
The temporal evolution of $A$, $q$, and $\Gamma$, (shown in Figure \ref{fig:IsoVStime}, \ref{Sfig:OddVStime} and \ref{Sfig:AnisoVStime}) were fitted with a double-exponential function convoluted with a Gaussian kernel, as previously utilized by \citet{unikandanunniAnisotropicUltrafastSpin2021} and \citet{zhouhagstromSymmetrydependentUltrafastManipulation2022}. Percent shift obtained using double exponential fits for $A_R$, $A_L$, $q_R$, $q_L$, $\Gamma_R$, and $\Gamma_L$ are shown in Table \ref{Stab:Amp_table}. Please note that double exponential fits could not be performed reliably for $q_L$ and $\Gamma_L$, so the results presented in Table \ref{Stab:Amp_table} are the cuts at maximum quench. Figure \ref{Sfig:timescales_and_fit}(a) shows both the data and the temporal fit for $A_R$. The residual ($\approx$~25 times smaller than the signal) is plotted on the top panel of the same figure. This temporal fit method was used to extract $A_R$, $A_L$, $\Delta q_R$ and $\Delta \Gamma_R$ plotted in figure \ref{fig:Aq_vs_fluence}. Quench time $\tau_m$ and recovery time $\tau_{rec}$ plotted in figure \ref{fig:t_min_vs_fluence} were also extracted using these temporal fits. Figure \ref{Sfig:timescales_and_fit}(b) and its inset, compares the temporal evolution data for $A_R$, $q_R$, and $\Gamma_R$, as extracted from the 2D fitting for the maximum pump-fluence of 13.4 mJ/cm$^2$. There is a visible difference in the apparent quench time and recovery time between $A_R$ and both $q_R$ and $\Gamma_R$ resulting in differences in $\tau_m$ and $\tau_{rec}$ extracted using the temporal fitting.

\begin{table*}[ht!]
\caption{Normalized percent change for amplitude ($A_R$, $A_L$), peak position ($q_R$, $q_L$) and peak width ($\Gamma_R$, $\Gamma_L$) for labyrinthine (subscript \emph{R}) and stripe (subscript \emph{L}) for various pump fluences. The presented results are the same as plotted in Fig. \ref{fig:Aq_vs_fluence} of the manuscript and were obtained using double exponential fits. Please note that double exponential fits could not be performed reliably for $q_L$ and $\Gamma_L$, so the results presented here are the cuts at maximum quench.}
\label{Stab:Amp_table}
\begin{tabular}{@{}ccccccccccccc@{}}
\toprule
                    &  & \multicolumn{5}{c}{Labyrinthine/Ring}                   &  & \multicolumn{5}{c}{Stripe/Lobes}                         \\ \cmidrule(r){1-1} \cmidrule(lr){3-7} \cmidrule(l){9-13} 
Fluence (mJ/cm$^2$) &  & $A_R (\%)$           &  & $q_R (\%)$          &  & $\Gamma_R (\%)$          &  & $A_L (\%)$            &  & $q_L (\%)$          &  & $\Gamma_L (\%)$          \\ \cmidrule(r){1-1} \cmidrule(lr){3-3} \cmidrule(lr){5-5} \cmidrule(lr){7-7} \cmidrule(lr){9-9} \cmidrule(lr){11-11} \cmidrule(l){13-13} 
0.8                 &  & $-7.15\pm4.12$  &  & $-0.07\pm0.44$ &  & $0.25\pm0.47$  &  & $-7.64\pm3.33$   &  & $0.05\pm0.31$  &  & $2.10\pm1.57$  \\
2.1                 &  & $-14.32\pm0.94$ &  & $-0.42\pm0.32$ &  & $0.79\pm0.87$  &  & $-14.78\pm1.49$  &  & $-0.39\pm0.27$ &  & $-2.51\pm1.37$ \\
5.0                 &  & $-17.41\pm3.06$ &  & $-1.04\pm0.96$ &  & $2.44\pm0.66$  &  & $-17.85\pm3.43$  &  & $0.37\pm0.36$  &  & $-3.98\pm2.42$ \\
5.9                 &  & $-19.16\pm6.05$ &  & $-1.34\pm1.20$ &  & $1.11\pm3.37$  &  & $-22.79\pm4.31$  &  & $0.47\pm0.73$  &  & $-2.49\pm4.67$ \\
6.7                 &  & $-21.07\pm1.86$ &  & $-1.31\pm2.66$ &  & $1.97\pm1.76$  &  & $-24.90\pm7.93$  &  & $0.23\pm0.28$  &  & $-2.94\pm2.07$ \\
7.5                 &  & $-23.01\pm3.09$ &  & $-1.17\pm0.81$ &  & $3.15\pm0.78$  &  & $-26.91\pm8.79$  &  & $-0.41\pm0.31$ &  & $-3.77\pm2.22$ \\
8.4                 &  & $-25.77\pm2.36$ &  & $-1.70\pm1.57$ &  & $4.19\pm4.03$  &  & $-30.64\pm8.52$  &  & $-0.82\pm0.47$ &  & $-4.14\pm3.72$ \\
9.2                 &  & $-27.82\pm2.27$ &  & $-2.29\pm1.31$ &  & $7.28\pm2.88$  &  & $-31.45\pm5.87$  &  & $-1.13\pm0.36$ &  & $-4.50\pm3.14$ \\
10.0                &  & $-32.47\pm2.35$ &  & $-2.81\pm0.70$ &  & $13.46\pm3.45$ &  & $-33.53\pm11.93$ &  & $-0.96\pm0.45$ &  & $-1.97\pm4.45$ \\
10.9                &  & $-35.18\pm3.53$ &  & $-3.31\pm0.90$ &  & $15.73\pm3.08$ &  & $-38.08\pm6.48$  &  & $-0.82\pm0.33$ &  & $-8.13\pm3.85$ \\
11.7                &  & $-34.33\pm2.90$ &  & $-3.73\pm0.94$ &  & $17.54\pm2.48$ &  & $-37.61\pm3.10$  &  & $-0.45\pm0.28$ &  & $-7.85\pm2.96$ \\
12.6                &  & $-35.91\pm3.18$ &  & $-4.32\pm0.82$ &  & $19.99\pm3.43$ &  & $-39.92\pm7.97$  &  & $-0.93\pm0.18$ &  & $-6.16\pm2.08$ \\
13.4                &  & $-38.51\pm1.82$ &  & $-5.29\pm0.80$ &  & $26.69\pm3.83$ &  & $-42.66\pm6.04$  &  & $-0.77\pm0.31$ &  & $-4.33\pm1.67$ \\ \bottomrule
\end{tabular}
\end{table*}

The extracted time-constants $\tau_m$ and $\tau_{rec}$ are significantly faster than those previously reported in \citet{zusinUltrafastPerturbationMagnetic2022}. This earlier study reported $\tau_m\approx$ 1 ps, and a subsequent recovery time $\tau_{rec}\approx$ 10 ps. While the magnitude of the distortions in \citet{zusinUltrafastPerturbationMagnetic2022} is approximately the same as reported here, with $\Delta M\approx$ 40\% and $\Delta q\approx$ 5\%, the difference in speeds is substantial. There are several significant differences in these two experiments that can account for these differences in response time: (1) Sample structure and thickness: In \citet{zusinUltrafastPerturbationMagnetic2022}, the sample is a 40-nm magnetic multilayer deposited on a 100 nm Si$_3$N$_4$ membrane. In contrast, the sample for this study is a 13-nm magnetic multilayer grown on a 100 nm Si membrane. Given an optical penetration length of only 7 nm for a Ni thin film \cite{axelevitchInvestigationOpticalTransmission2012}, we expect that the vertical profile of magnetic quenching for our film is more substantial than that in \citet{zusinUltrafastPerturbationMagnetic2022}, which was shown by modeling to only reach approximately 10 nm into the depth of the 40-nm sample. In addition, it is plausible that the mechanics of domain wall movement induced by ultrafast pumping would be quantitatively different due to micromagnetic differences between samples of such varying thicknesses. (2) Pump fluence: \citet{zusinUltrafastPerturbationMagnetic2022} reports a damage-threshold-limited measurement for a single pump fluence of 23 mJ/cm$^2$. The maximum pump fluence for this study, also limited by the threshold for sample damage, is 13.4 mJ/cm$^2$. (3) Domain pattern: The domain structure reported in \citet{zusinUltrafastPerturbationMagnetic2022} was purely labyrinthine, whereas the sample used in this study is a linearly-textured labyrinth pattern. The admixture of labyrinthine and stripe domains resulted from the exposure to the optical pump beam over the course of ~15 minutes, similar to what was observed in \citet{zhouhagstromSymmetrydependentUltrafastManipulation2022}. (4) Time resolution: The requirement of high dynamic range for the measurements in \citet{zusinUltrafastPerturbationMagnetic2022} limited the time-resolution to 400 fs. This prevented  detection of any ultrafast components of the initial magnetization response as shown in Fig. \ref{fig:t_min_vs_fluence}, which were all on time scales less than 400 fs. 

\begin{figure}[t]
    \centering
    \includegraphics[width = 0.7\columnwidth]{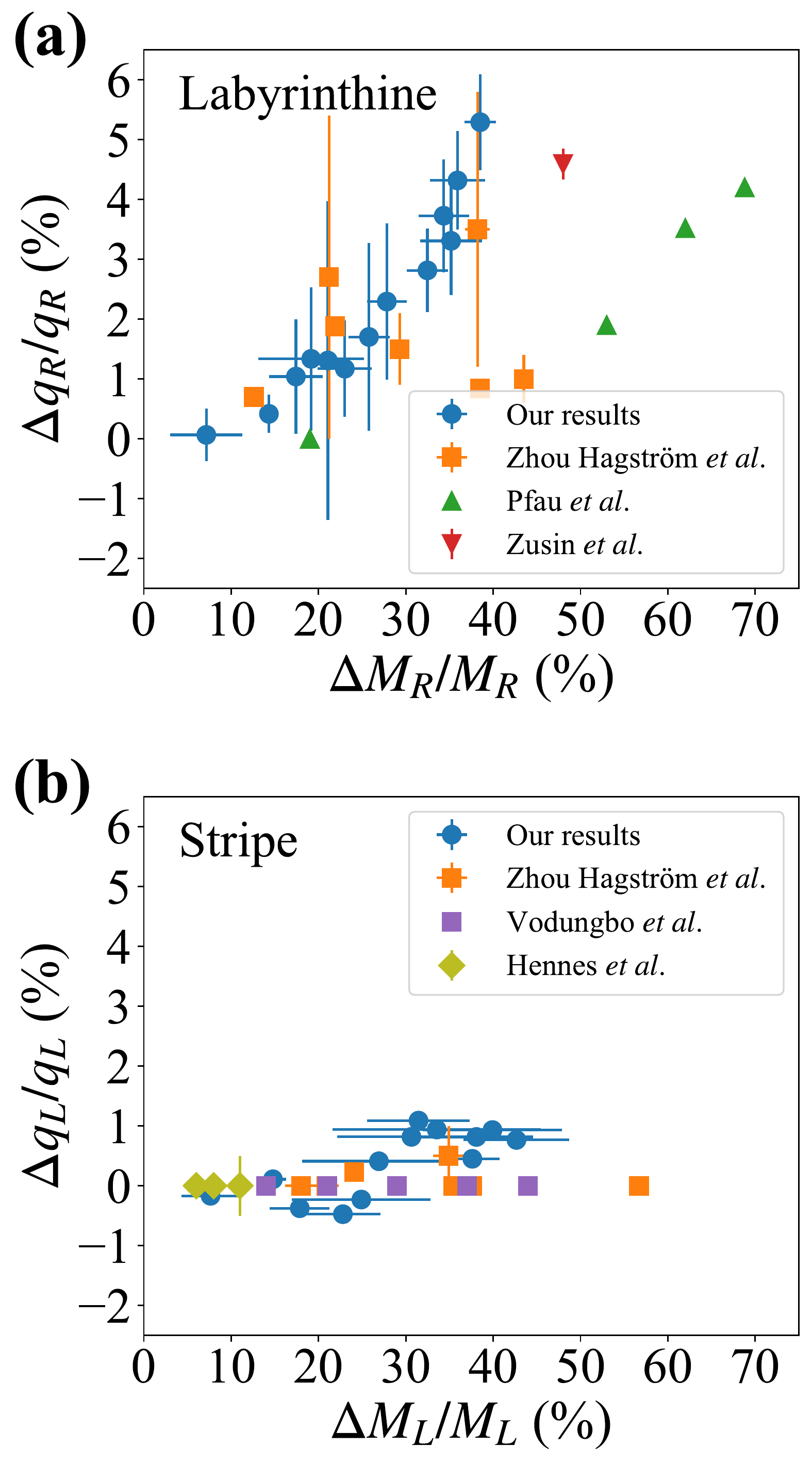}
    \caption{\textbf{Radial shift vs Magnetization quench}. The figure compares the normalized change in radial peak position as a function of normalized magnetization change from this study with other publications for labyrinth \textbf{(a)} and stripe \textbf{(b)} domains~\citet{zhouhagstromSymmetrydependentUltrafastManipulation2022, pfauUltrafastOpticalDemagnetization2012, zusinUltrafastPerturbationMagnetic2022, vodungboLaserinducedUltrafastDemagnetization2012, hennesLaserinducedUltrafastDemagnetization2020}.}
    \label{Sfig:dqvsdM_labvs_stripe}
\end{figure}

\subsection{\label{Comparison with lit} \texorpdfstring{$\Delta q/q_0$ vs $\Delta M$}{Lg} comparison with the literature}
Fig. \ref{Sfig:dqvsdM_labvs_stripe} compares our results to previously reported ultrafast measurements of magnetic domain pattern evolution. Fig. \ref{Sfig:dqvsdM_labvs_stripe}(a) plots the normalized peak shift ($\Delta q/q_0$) as a function of magnetization quench ($\Delta M/M_0$) for labyrinth domains by \citet{zhouhagstromSymmetrydependentUltrafastManipulation2022}, \citet{pfauUltrafastOpticalDemagnetization2012}, \citet{zusinUltrafastPerturbationMagnetic2022} and this study. Our measurements (blue circles) are in good agreement with the measurements done at the European XFEL on the same sample \cite{zhouhagstromSymmetrydependentUltrafastManipulation2022}. Furthermore, our measurements also indicate the presence of a similar threshold present in previous studies where no peak shift is observed for low fluence. The comparison of $\Delta q$ for different studies shows that these effects are ubiquitous for all samples that support labyrinthine domains, although the dependence on fluence is strongly material dependent, such as the multilayer stack design, pinning sites, defect density, and materials used. Fig. \ref{Sfig:dqvsdM_labvs_stripe}(b) shows $\Delta q$ for stripe-like domains as a function of pump fluence in \citet{zhouhagstromSymmetrydependentUltrafastManipulation2022}, \citet{vodungboLaserinducedUltrafastDemagnetization2012} and \citet{hennesLaserinducedUltrafastDemagnetization2020}, as well as this study. Only marginal or no shifts in $q$ were observed in all four studies which is consistent with minimal to no shift observed for stripe domains. 

\subsection{\label{Micromagnetic simulations}Micromagnetic modeling}
We used micromagnetic simulations to investigate domain-wall distortions due to changes in curvature and its effect on the EUV scattering. MuMax3~\cite{vansteenkisteDesignVerificationMuMax32014} was used for the simulations and ran in a GTX QUADRO 5000 GPU. In order to model $\Delta q$, we simulated the initial magnetization distribution for a sample with parameters similar to that used in this study. The magnetization was allowed to relax into an equilibrium labyrinthine domain-pattern. The saturation magnetization was then decreased instantaneously by 40\% of its original value similar to the experimentally observed magnetization quench (Fig. \ref{fig:Aq_vs_fluence}) for the highest pump fluence of 13.4 mJ/cm$^2$. The system was then allowed to relax to a transient intermediate state at 200 ps.

We use magnetization parameters consistent with the sample: saturation magnetization $M_s=771$~kA/m, exchange constant $A=20$~pJ/m, first-order uniaxial anisotropy $K_u=739$~kJ/m$^3$ and second-order uniaxial anisotropy $K_{u2}=-266$~kJ/m$^3$. The labyrinthine domain pattern is stabilized from a random initial condition while an experimental data set for a similar magnetic system was used for the initial condition for the stripe domain pattern.

\begin{figure*}[ht]
    \centering
    \includegraphics[width = \textwidth]{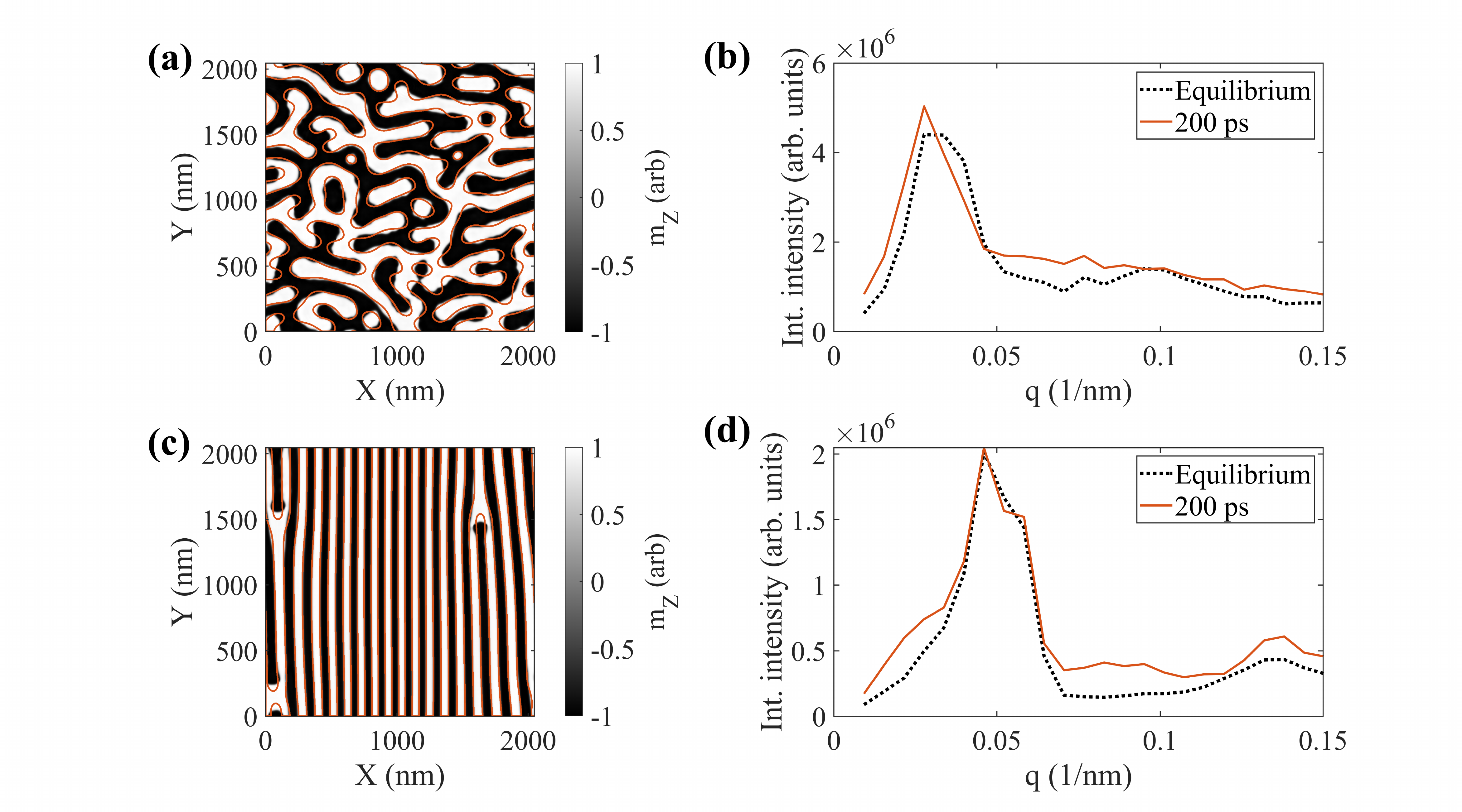}
    \caption{\textbf{Micromagnetic simulations of domain pattern}. Modified simulated (black and white) and initial domain pattern (red outline) for \textbf{(a)} labyrinthine and \textbf{(c)} stripe domain pattern. The modified simulated pattern was obtained by reducing the magnetization by 40 \% for both labyrinth and stripe domains respectively. A decrease in the curvature of the curved domain wall is observed for both domain patterns. Azimuthally integrated intensity from a 2D fast Fourier transform for \textbf{(b)} labyrinthine and \textbf{(d)} stripe domain pattern. The dashed black line is for the initial state and the solid red line is for the modified state at 200 ps. A clear shift in the ring radius is observed for labyrinthine domains whereas no shift is observed for stripe domains.}
    \label{Sfig:sim_results}
\end{figure*}

The simulated modified domain pattern at 200 ps for both labyrinthine and stripe domains is presented in Figure \ref{Sfig:sim_results}(a) and (c), respectively. The red outline traces the domain boundaries for the initial equilibrium state. The displacement of the curved domain walls is observed for both labyrinths and stripes, but the effect is more pronounced in the labyrinthine domain pattern due to the abundance of curved walls. For stripes, the most prominent displacements occurred for cap walls where the stripe domains terminate. Note that while the time scale for the wall displacement is not accurate due to the use of micromagnetic simulations \cite{iacoccaSpincurrentmediatedRapidMagnon2019},  the dependence of wall displacement on wall curvature allows us to examine how such domain rearrangement affects the diffraction pattern in the reciprocal space. 

To simulate X-ray scattering, an FFT was applied to both the initial and modified domain patterns. In the case of the labyrinthine domains, the FFT yielded an isotropic ring in reciprocal space, in qualitative agreement with the experimental results. The azimuthal integral (in reciprocal space) of the ring is shown in Figure \ref{Sfig:sim_results} \textbf{(b)}. Both a reduction of the ring radius and a broadening of the ring width occur due to domain rearrangement at 200 ps. The initial ring radius is 0.0329 nm$^{-1}$, similar to the experimental initial ring radius of 0.0366 nm$^{-1}$. The final ring radius is 0.0299 nm$^{-1}$. The initial ring width is 0.0237 nm$^{-1}$, significantly broader than the experimental ring width of 0.0130 nm$^{-1}$. This is attributed to the higher degree of disorder for the micromagnetic simulation. 

In spite of the high degree of disorder in the micromagnetic domain pattern, the ring width still broadens as a result of the wall displacements, with a final value of 0.0254 nm$^{-1}$. The ring broadening relative to the radial decrease is smaller than the experiment. Again, this is because the simulated domain pattern is more disordered than the experimental domains. The FFT for the stripe domains does not yield any discernable shift in radius or width after relaxation for the reduced saturation magnetization. This is attributed to the low density of curved segments of the domain-walls. Note that if the micromagnetics simulations were allowed to run longer there would be increase in the domain periodicity due to lowering of saturation magnetization. However, such homogenous changes in the domain periodicity at ultrafast timescales would lead to extremely fast expansion of the domain widths, which would be unphysical \cite{pfauUltrafastOpticalDemagnetization2012}.

The explanation why a reduction in the saturation magnetization results in the wall displacement, we mapped a curved domain wall into one dimension by defining a profile dependent on the total moment $\mu(x) = \mu_a\tanh{\frac{x}{a}}+\mu_{av}$, where $\mu_a$ is the asymptotic value of a symmetric domain profile, and $\mu_{av}=\mu(x=0)$. Under the assumption that ultrafast quenching minimizes the non-local dipole field that stabilizes the texture in equilibrium, we seek a new symmetric distribution for $\mu(x)$. This leads to a net shift of around $\mathrm{arctanh}\left(\frac{a\kappa}{2}\right)$, where $\kappa$ is the domain curvature. This simple argument explains why curved domains are more prone to motion based on exchange energy, but as the model is based on micromagnetics it does not capture ultrafast motion.

\subsection{\label{Velocity Calculation}Domain wall velocity calculation}
Numerous localized regions with shift in domain walls can be identified in the simulated domain pattern images (\ref{Sfig:sim_results}(a)). These localized shifts are predominantly located in regions of significant domain wall curvature with a common characteristic that the shift tends to reduce the wall curvature. Given the inherent randomness of the labyrinthine domain structure, the wall shift will induce local changes in the domain area which affects both black and white domains. Statistically, these localized displacements of the domain wall area should average to zero, i.e. neither the black or white domains increase in area at the expense of the other. However, the root mean square (rms) of the displacement of the domain area will be nonzero and was used to estimate the domain wall velocity. Using image analysis on the simulated domain pattern, we measured the rms area of the localized displacement to be 2100 nm$^2$. Using the curvature density of 75.1 $\mu m^{-2}$ we can express the localized displacement as an areal fraction of the average domain size, we obtain $\Delta A_{RMS}/A = 2100\times75.1\times10^{-6} = 0.16$.

The fractional change in the ring radius in Fig. \ref{Sfig:sim_results}(b) is $(\Delta q/q)_{sim}$ = 0.095. The ratio of the fractional areal changes to fractional radial change is  $K = \left(\Delta A_{RMS}/A\right)/\left(\Delta q/q\right)_{sim} \approx$ 1.65. For the rest of the analysis, we assume that this is a proportionality constant between the fractional areal change at the curved sites in real space and the fractional radial change of the ring radius in reciprocal space. Note that this proportionality may depend on the details of the labyrinthine geometry. However, any such dependence should be weak because the general randomness inherent in all the meandering labyrinthine structures precludes any coherent scattering effects that might otherwise have a strong effect on the proportionality. Furthermore, we performed five different micromagnetic simulations with a change in saturation magnetization ranging from 10 \% to 40 \%, all of which showed that $K$ varies in a narrow range from 1.65 to 1.9. An average value of $K$ = (1.84 $\pm$ 0.11) was used for velocity calculations.

The domain wall velocity was calculated using the following equation, 
\begin{equation}
    v = K\frac{\left(\Delta q/q\right)_{exp}}{\rho_{c} w \tau_m}.
    \label{SEqn:vel_eqn}
\end{equation}

Here $(\Delta q/q)_{exp}$ is the experimentally observed fractional change in diffraction ring radius and $\tau_m$ is the time constant for the radial shift obtained from temporal fits to the experimental data. $K$ is the proportionality constant as defined above. $w$ is the average domain width and $\rho_{c}$ is the curvature density obtained from the MFM images. The maximum experimental fractional change in the diffraction ring radius $(\Delta q/q)_{exp}$ at the maximum fluence of 13.4 mJ/cm$^2$ is (0.0555 $\pm$ 0.001) $nm^{-1}$. Curvature density ($\rho_{c}$) in the MFM image was estimated to be (60.39 $\pm$ 6.26) $\mu m^{-2}$ by fitting curvatures post edge detection in MATLAB. Uncertainty in curvature density was calculated using the disparity between the density for dark (down) and light (up) contrast domains in MFM. An average domain width ($w = \pi/q_0$) of 85.8 nm was calculated using the values of $q_R$ for $t<0$ from the 2D fit of the magnetic scattering. Using the time constant for the radial shifts ($\tau_m\approx$ 0.30 $\pm$ 0.08) ps at the maximum fluence, the effective maximum speed of the wall displacement is (66 $\pm$ 20) km/s. 

\bibliographystyle{apsrev4-2}
\bibliography{references}

\end{document}